\documentclass{moriond}

\def\be{\begin{equation}}
\def\ee{\end{equation}}
\def\bea{\begin{eqnarray}}
\def\eea{\end{eqnarray}}

\newcommand{\Photo}{}

\begin{document}
\vspace*{4cm}
\title{Searching for new physics during gravitational waves propagation}

\author{Leïla Haegel}

\address{Université de Paris, CNRS, AstroParticule et Cosmologie, F-75013 Paris, France}

\maketitle\abstracts{
The direct detection of gravitational waves by ground-based interferometers opened an unprecedented channel to probe alternative theories of gravitation. 
Several theories predict a dispersion of the gravitational waves during their propagation, distorting the signals observed by LIGO and Virgo compared to their predictions from general relativity.
Such dispersion could induce a modification of the luminosity distance inferred with gravitational radiation with regards to electromagnetic radiation.
By analysing two multimessenger events, we set constraints on a large class of proposed theories, including extra-dimensional and scalar-tensor theories.
The multimessenger events are the binary neutron star merger \textit{GW170817} associated to \textit{\textit{GRB170817A}}, and the binary black hole merger \textit{ \textit{GW190521}} with postulated candidate electromagnetic counterpart \textit{\textit{ZTF19abanrhr}}. 
Without relying on multimessenger emission, a class of proposed theories predict a frequency-dependent dispersion of the gravitational waves breaking Lorentz invariance.
By analysing 31 GW events from binary-black holes coalescence, we constrain several coefficients parameterising Lorentz violation, including the best constraint on the graviton mass.
}

\section{Introduction}

Since the first direct detection of gravitational waves (GW) by the LIGO-Virgo collaboration (LVC) in 2015, 50 events have been reported during three observation runs \cite{Abbott:2020niy}. 
The GW observations originate from the coalescence of binary systems of compact objects, of which the LVC interferometers record the late inspiral, plunge and merger. 
Most of the compact objects are stellar-mass black holes, the other type being neutron stars characterised by their low mass and potential presence of tidal effects. 
The closest event detected is \textit{GW170817} \cite{TheLIGOScientific:2017qsa}., the coalescence of two binary neutron stars at $40^{+8}_{-14}$~Mpc, and the furthest event detected is \textit{GW190413\_134308}, the coalescence of two black holes at $4.45^{+2.48}_{-2.12}$~Gpc \cite{Abbott:2020niy}.

GW offer a new way to probe fundamental physics with purely gravitational signals. 
Several alternatives to General Relativity (GR), including scalar-tensor theories of gravity or effective field theories such as the Standard Model Extension, predict that GW may disperse during their propagation.
The dispersion can lead to a distortion of the pattern seen by the LVC interferometers and a modification of the apparent luminosity distance inferred from the amplitude of the GW signal.
Section~\ref{sec:friction} describes constraints on several parameterisation of alternative theories of gravitation using multimessenger events.
Section~\ref{sec:dispersion} describes the constraints on different coefficients responsible for Lorentz invariance violation obtained from GW signals only.

\section{Constraining GW friction with multimessenger events}
\label{sec:friction}

Multimessenger events consist of the simultaneous observation of emissions from different channels, including gravitational and electromagnetic (EM) radiations, as well as astroparticles. 
The coincident observation of GW and EM signals in ``standard sirens" events enable to measure the expansion of the Universe from the combined information of the EM redshift and the GW amplitude. 
The LVC has published the first measurement of the Hubble constant inferred from the redshift and luminosity distance of the event \textit{GW170817} described below \cite{LIGOScientific:2017adf}.
The source of this expansion being currently unknown, several classes of alternative theories of gravitation have been proposed to explain its origin.
A part of the phenomenology induced by those theories consists of GW friction, i.e. a modification of the wave amplitude due to dispersion leading to a different luminosity distance inferred from GW and EM signals, i.e. $d_L^{GW} \neq d_L^{EM}$.
By analysing several multimessenger events, we measure the Hubble constant $H_0$, the matter density $\Omega_{m,0}$, as well as the coefficients parameterising the difference in the luminosity distances inferred from GW and EM signals from several theories.

In this work, we analyse two events: 
\begin{itemize}
\item \textit{GW170817}, the coalescence of two neutron stars (kilonova event) at $40^{+8}_{-14}$~Mpc emitting both GW and the short $\gamma$-ray burst \textit{GRB170817A}, followed by EM emissions in a large spectrum of wavelength \cite{GBM:2017lvd}.
\item  \textit{GW190521}, the coalescence of two black holes at $3.92^{+2.19}_{-1.95}$~Gpc with GW detected by the LVC and the candidate EM counterpart \textit{ZTF19abanrhr} possibly due to the surrounding environment of an AGN disk \cite{Graham:2020gwr}. While the association of the two events has not been confirmed, we include them to study the impact of a large redshift and additional degrees of constraints on the analysis.
\end{itemize}

\begin{figure}[h!]
\begin{minipage}{0.5\linewidth}
\centerline{\includegraphics[width=0.9\linewidth]{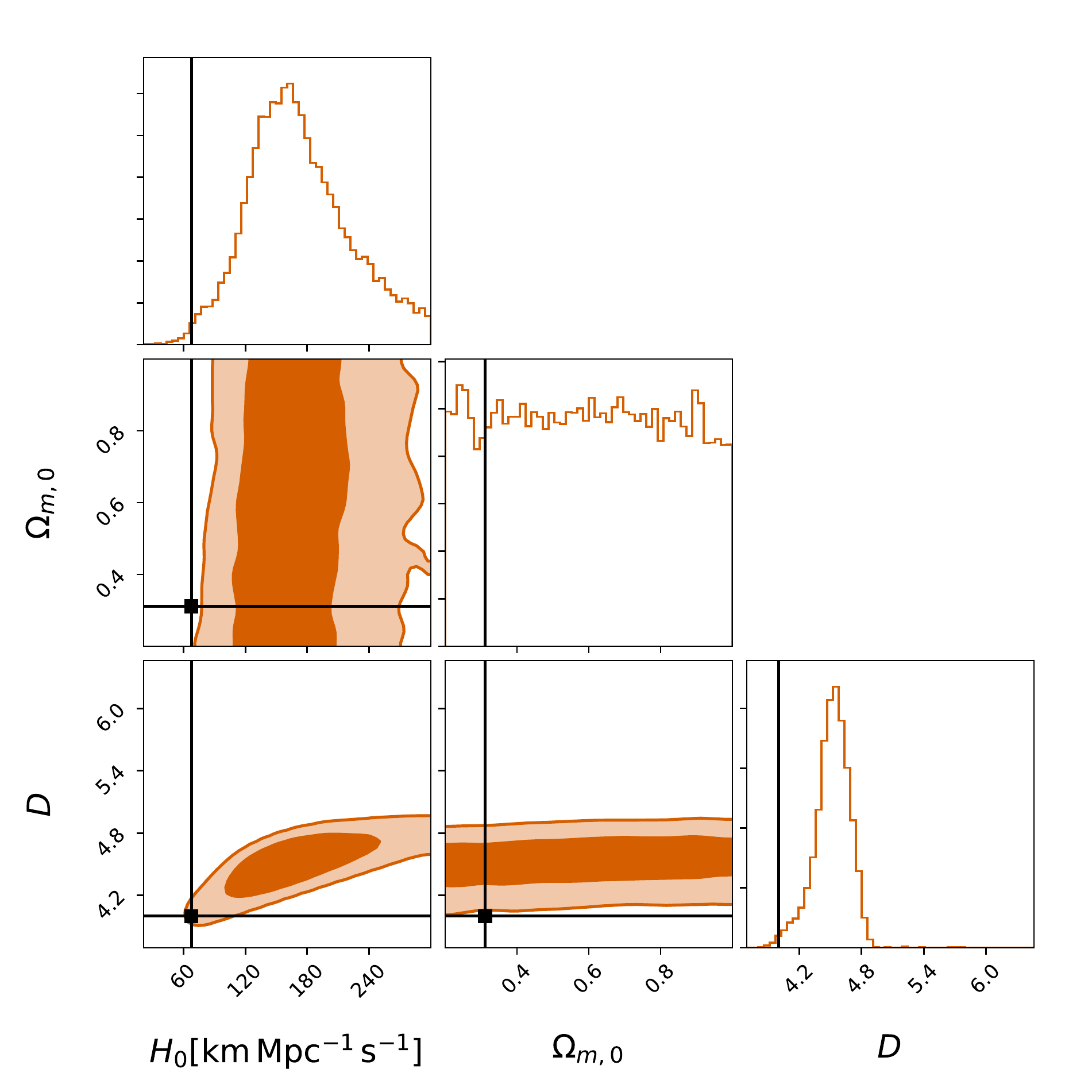}}
\end{minipage}
\hfill
\begin{minipage}{0.5\linewidth}
\centerline{\includegraphics[width=0.9\linewidth]{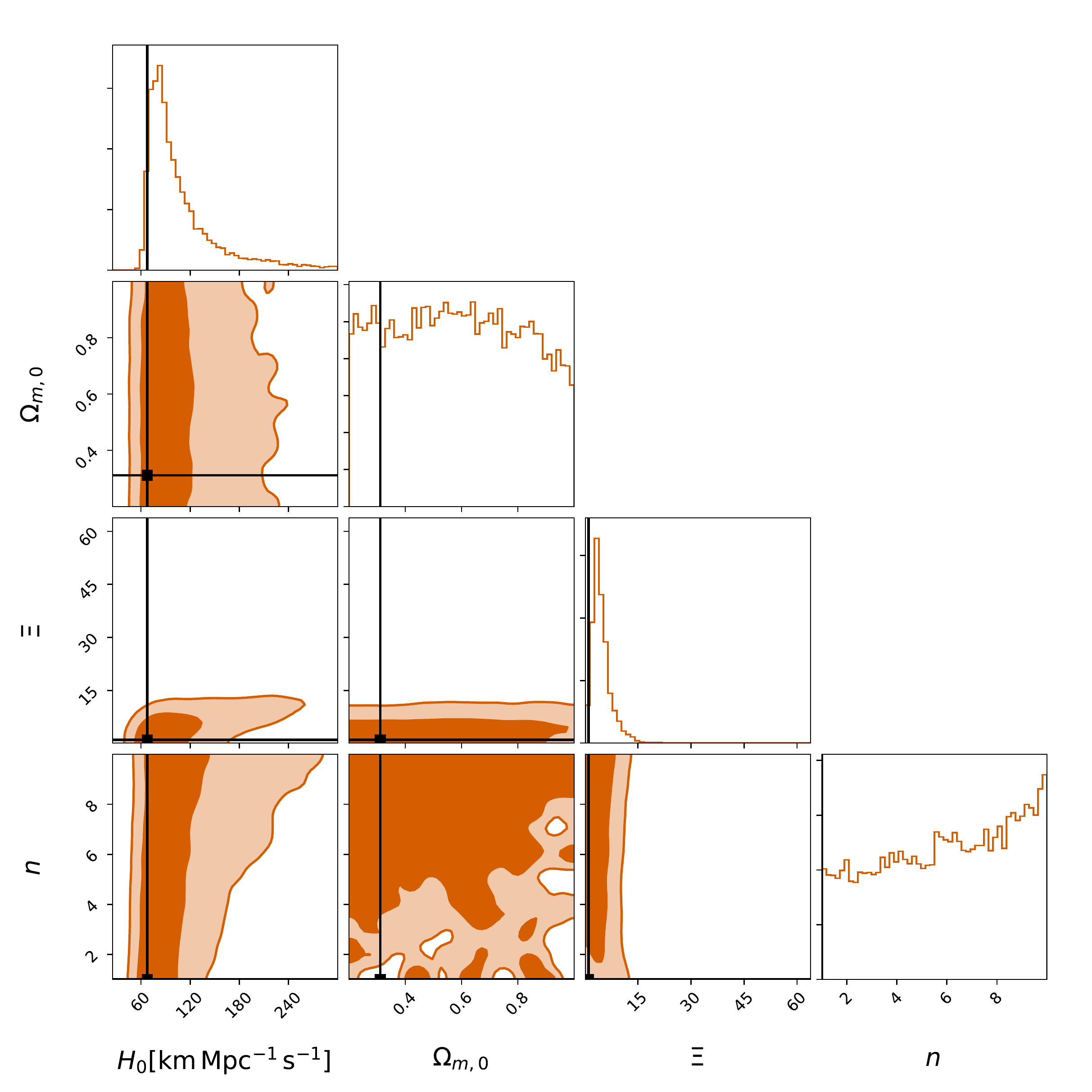}}
\end{minipage}
\hfill
\caption[]{Constraints on the Hubble constant $H_0$, matter density $\Omega_{m,0}$ and alternative theories of gravitation parameters \cite{Mastrogiovanni:2020mvm}. Left: constraints on the number of spacetime dimensions $D$. Right: constraints on the scalar-tensor theories coefficients $\Xi$ and $n$.}
\label{fig:gwfriction}
\end{figure}

\paragraph{Extra dimensions:}
Several alternative theories of gravitation, including proposals of quantum gravity \cite{Calcagni:2019kzo} and Dvali-Gabadadze- Porrati (DGP) gravity \cite{Dvali:2000hr}, assume a spacetime with more than 4 dimensions.
In the case where those extra dimensions are not compactified, they are characterised by the length scale $R_C$  beyond which gravitation is modified from GR.
The GW additional energy loss in those dimensions modifies the GW luminosity distance as: 
\begin{equation}
    d_L^{\rm GW} = \left[1+\left(\frac{d_L^{\rm EM}}{R_c}\right)^n \right] ^{\frac{D-2}{2n}},
\end{equation}
where $D$ is the number of spacetime dimensions. As shown on Figure~\ref{fig:gwfriction} left, the events are consistent with a 3+1 spacetime. 
 
\paragraph{Scalar-tensor theories:}

A large set of alternative theories of gravitation, including Brans-Dicke, Horndeski and 
Degenerate Higher-Order Scalar-Tensor (DHOST) theories, are constructed by adding a scalar field to the tensor field of GR. 
A subset of those theories predict a modification of the GW luminosity distance similar to the one predicted from non-local gravity \cite{Maggiore:2014sia}, that is: 
\begin{equation}
    d_L^{\rm GW} = d_L^{\rm EM}  \left[\Xi + \frac{1-\Xi}{(1+z)^n}\right],
\end{equation}
where $n>0$ is the spacetime stiffness and $\Xi>0$ recovers GR for $\Xi=1$.
As shown on Figure~\ref{fig:gwfriction} right, the observations are consistent with GR. 

\paragraph{Time-varying Planck mass:}

A time-varying Planck mass arises in several alternative theories of gravitation, and impacts the GW luminosity distance in case of the presence of an extra field impacting the expansion of the Universe as \cite{Lagos:2019kds}: 
\begin{equation}
d_L^{\rm GW} = d_L^{\rm {EM}} {\rm{exp}} \left[\frac{c_M}{2\Omega_{\Lambda,0}} \ln \frac{1+z}{\Omega_{m,0}(1+z)^3+\Omega_{\Lambda,0}} \right],
\end{equation}
 where $c_M$ parametrise the evolution of the dark energy content in the Universe. 
The analysis of the multimessenger events lead to a measurement consistent with GR where $c_M=0$. 

\section{Constraining Lorentz invariance violation with gravitational-waves observations}
\label{sec:dispersion}

The Standard Model Extension (SME) formalism is an effective field theory designed for phenomenological searches of new physics.
It has been developed to probe low-energy manifestation of a possible unified theory at the Planck scale, and has been extensively probed in the particle sector.
Recently, it has been extended to search for deviations from General Relativity, by adding new fields to the linearised GR Lagrangian \cite{Kostelecky:2016kfm}: 
\begin{equation}
\renewcommand{\arraystretch}{1.2}
\begin{array}{rc@{\,}c@{\,}l}

\mathcal{L} & = &&  \frac{1}{4} \varepsilon^{\mu \rho \alpha \kappa} \varepsilon^{\nu \sigma \beta \lambda} \eta_{\kappa \alpha} h_{\mu \nu} \delta_{\alpha} \delta_{\beta} h_{\rho \sigma}  \\
       &   & + & \frac{1}{4} h_{\mu \nu} (\hat{s}^{\mu \rho \nu \sigma} + \hat{q}^{\mu \rho \nu \sigma} + \hat{k}^{\mu \rho \nu \sigma}) h_{\rho \sigma} ,\\ 
\end{array}
\label{eq:spa}
\end{equation}

where the first line is the linearised Einstein-Hilbert Lagrangian, and the second line contains three
gauge-invariant irreducible operators that can be described in 3 classes of increasing minimal mass dimensions, starting from $d=4$.

The presence of new fields leads to a frequency-dependent modification of the speed of the GW, inducing dispersion and a breaking of Lorentz symmetry \cite{Tasson:2021loi}.
Dispersion effects arise for $d \geq 5$, while $d=4$ coefficients lead to dispersion-free propagation. 
The dispersion is polarisation-dependent for a certain class of coefficients, inducing CPT-even and CPT-odd birefringence according to the mass dimension parity \cite{Mewes:2019dhj}.

\paragraph{Polarisation-independent dispersion:}

A parameterisation of the dispersion can be written as a modification of the energy conservation relation \cite{Mirshekari:2011yq}:
\begin{equation}
E^2 = p^2 c^2 + A_{\alpha} p^{\alpha} c^{\alpha},
\label{eq:disp}
\end{equation}
where $A_{\alpha}=0$ corresponds to GR, and integer $A_{\alpha}$ can be associated to SME coefficients.

\begin{figure}[h!]
\begin{minipage}{0.9\linewidth}
\centerline{\includegraphics[width=0.6\linewidth]{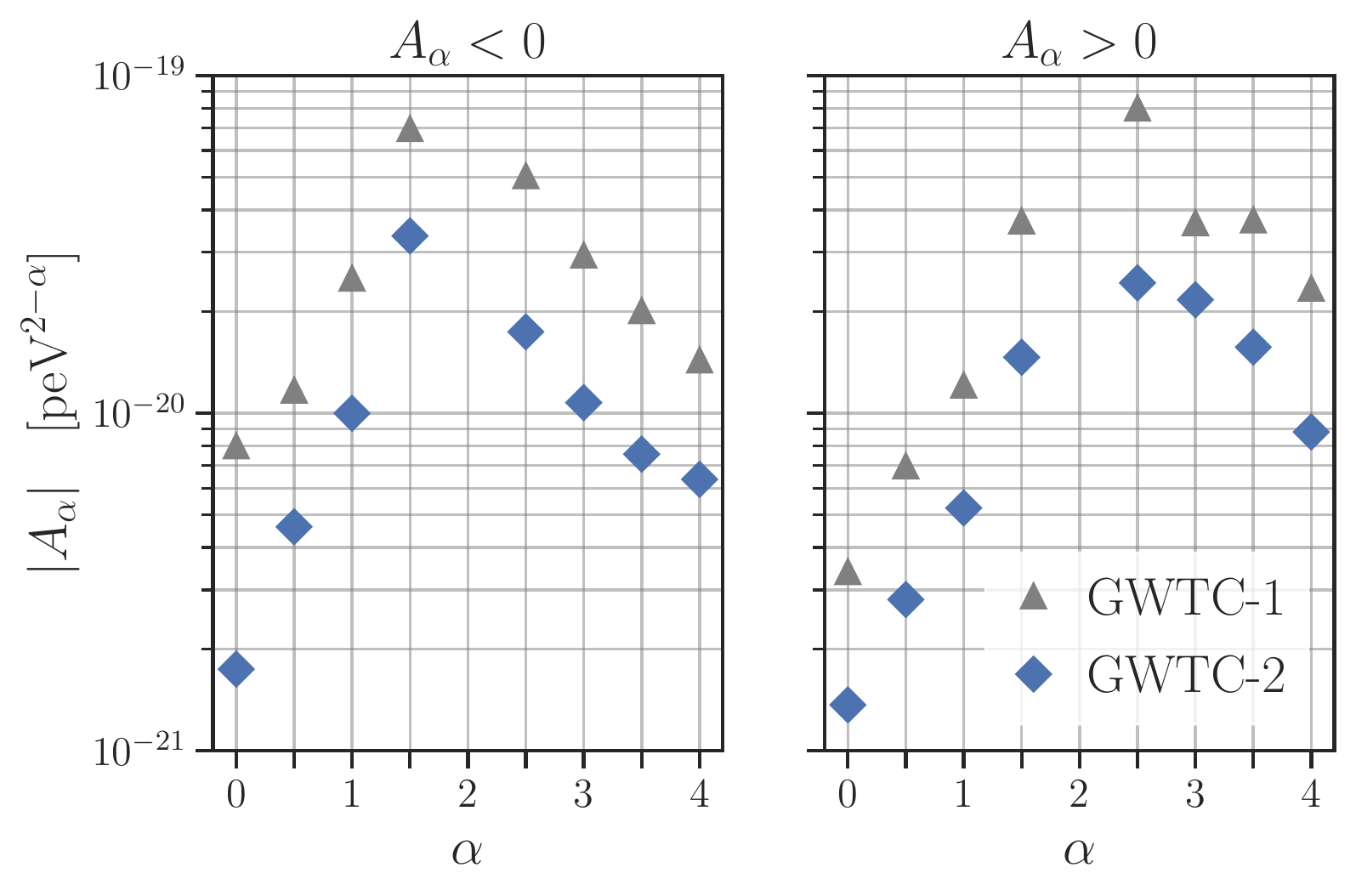}}
\end{minipage}
\caption[]{Constraints on the dispersion coefficients $A_{\alpha}$ of Eq.(\ref{eq:disp}). The blue diamond and grey triangle correspond to the different integrated catalog of GW event detected by the LVC \cite{Abbott:2020jks}.}
\label{fig:gwdispersion}
\end{figure}

The dephasing of the GW signal induced by $A_{\alpha}$ has been probed for incremental values of 0.5 for $\alpha \in [0,4]$, with the exception of $\alpha=2$ that is degenerate with the coalescence time. 
Correspondancies can be made between several values of $\alpha$ and alternative theories of gravity, including massive gravity for $A_{\alpha=0} >0$, multi-fractal spacetime for $A_{\alpha=2.5}$, doubly special relativity for $A_{\alpha=3}$, Ho\v{r}ava-Lifshitz and extra dimensional theories $A_{\alpha=4}$ \cite{LIGOScientific:2019fpa}.
The analysis of the GW from 31 binary black-hole coalescence events detected during the three first observational runs of the LVC leads to the constraints shown on Figure~\ref{fig:gwdispersion}. 
The corresponding constraint on the graviton mass is $m_g \leq 1.76 \times 10^{-23}$ eV / c$^2$.

\paragraph{Polarisation-dependent dispersion:}

In the SME, starting $d=5$ and for increasing odd values of the mass dimension, the dispersion impacts the $h_+$ and $h_{\times}$ polarisations of the GW differently.
The modification of the GW signal can be computed analytically and is function of the $\hat{k}$ operator, the GW spherical harmonics $Y_{lm}$, the redshift $z$ and the Hubble constant $H(z)$ \cite{Mewes:2019dhj}:

\begin{equation}
\renewcommand{\arraystretch}{1.2}
\begin{array}{rc@{\,}c@{\,}l}

h_{+} & = &&   f(\hat{k}, Y_{lm}, z, H(z)) h_{+}^{GR} + g(\hat{k}, Y_{lm}, z, H(z)) h_{\times}^{GR}  \\
h_{\times} & = &&   f'(\hat{k}, Y_{lm}, z, H(z)) h_{+}^{GR} + g'(\hat{k}, Y_{lm}, z, H(z)) h_{\times}^{GR}  \\
\end{array}
\label{eq:biref}
\end{equation}

Eq.(\ref{eq:biref}) lead to a different speed of propagation for the two GW polarisations, inducing a birefringent nature of the spacetime medium. 

\section*{Acknowledgments}

LH is supported by the Postdoc Mobility Grant 199307 of the Swiss National Science Foundation.
The author(s) would like to acknowledge the contribution of the COST Actions CA16104 and CA18108.
The authors gratefully acknowledge the support of the United States National Science Foundation (NSF) for the construction and operation of the LIGO Laboratory and Advanced LIGO as well as the Science and Technology Facilities Council (STFC) of the United Kingdom, the Max-Planck-Society (MPS), and the State of
Niedersachsen/Germany for support of the construction of Advanced LIGO  and construction and operation of the GEO600 detector.  Additional support for Advanced LIGO was provided by the Australian Research Council. The authors gratefully acknowledge the Italian Istituto Nazionale di Fisica Nucleare (INFN),  
the French Centre National de la Recherche Scientifique (CNRS) and the Netherlands Organization for Scientific Research,  for the construction and operation of the Virgo detector and the creation and support  of the EGO consortium.
The authors gratefully acknowledge the support of the NSF, STFC, INFN and CNRS for provision of computational resources.

\end{document}